\documentclass[aps,showpacs,twocolumn,superscriptaddress]{revtex4-1}
\usepackage{graphicx}
\usepackage{amsfonts}
\usepackage{amssymb}
\usepackage{amsmath}
\usepackage{natbib}
\usepackage{dcolumn}
\usepackage{bm}
\begin{document}
\title{$\Delta$ isobars in hyperon stars in a modified quark meson coupling 
model}
\author{H.S. Sahoo}
\author{R.N. Mishra}
\affiliation{Department of Physics, Ravenshaw University, Cuttack-753 003,
India}
\author{P.K. Panda}
\author{N. Barik}
\affiliation{Department of Physics, Utkal University, Bhubaneswar-751 004,
India}
\begin{abstract}
The possibility of the appearance of $\Delta$ isobars in neutron star matter 
and the so called {\it $\Delta$ puzzle} is studied in a modified quark meson 
coupling model where the confining 
interaction for quarks inside a baryon is represented by a phenomenological 
average potential in an equally mixed scalar-vector harmonic form. 
The hadron-hadron interaction in nuclear matter is then realized by 
introducing additional quark couplings to $\sigma$, $\omega$, and $\rho$ 
mesons through mean-field approximations. The couplings of the $\Delta$ to the 
meson fields are fixed from available constraints while the hyperon couplings 
are fixed from the optical potential values. It is observed that within the 
constraints of the mass of the precisely measured massive pulsars, 
PSR J0348+0432 and PSR J1614-2230, neutron stars with a composition of 
both $\Delta$ isobars and hyperons is possible. 
It is also observed that with an increase in the vector coupling strength of 
the $\Delta$ isobars there is a decrease in the radius of the neutron stars.  
\end{abstract}
\maketitle

\section{Introduction}
The investigations pertaining to the formation of baryons heavier than 
the nucleon at the core of neutron stars and the effects of such formation 
on the mass and radius of neutron stars is a subject of active research in 
nuclear astrophysics. It is expected that high density nuclear matter may 
consist not only of nucleons and leptons but also several exotic components 
such as hyperons, mesons as well as quark matter in different forms and 
phases. While many studies have been conducted to address the appearance of 
hyperons and on the so called {\it hyperon puzzle} \cite{debarati}, 
little work has been done to study the appearance of $\Delta$ $(1232)$ 
isobars in neutron stars. 
An earlier work \cite{glend} indicated the appearance of $\Delta$ at much 
higher 
densities than the typical densities of the core of neutron stars and hence was 
considered of little significance to astrophysical 
studies. However, recent studies \cite{drago,angli} suggest the possibility 
of an early appearance of $\Delta$ isobars with consequent softening of the 
equation of state (EOS) of dense matter. This leads to a reduction in the 
maximum mass of neutron stars below the current observational limit of 
$2.01\pm0.04$~M$_{\odot}$ \cite{antoniadis}.    

In the present work, we would like to address the $\Delta$-puzzle in a modified 
quark-meson  coupling model (MQMC)
\cite{rnm,hss} which has already been adopted successfully studying various bulk
properties of symmetric and asymmetric nuclear matter.  
The MQMC model is based on a confining 
relativistic independent quark potential model rather 
than a bag to describe the baryon structure in vacuum. 
The baryon-baryon interactions are realized by 
making additional quark couplings to $\sigma$, $\omega$, and $\rho$ mesons 
through mean-field approximations. More 
recently \cite{hss2} it was extended to study the EOS of nuclear 
matter with the inclusion of hyperons as new degrees of freedom and 
the effect of a 
non-linear $\omega$-$\rho$ term keeping in view the constraints of
the mass of the precisely measured massive pulsars, 
PSR J0348+0432 and PSR J1614-2230 .

Here, we include the delta isobars 
($\Delta^-$, $\Delta^0$, $\Delta^+$, $\Delta^{++}$) 
together with hyperons as new degrees of freedom in dense hadronic matter 
relevant for neutron stars. The interactions between nucleons, $\Delta$'s 
and hyperons in dense matter is studied and the possibility of the existence 
of the $\Delta$ baryon at densities relevant to neutron star core as well as 
its effects on the mass of the neutron star is analysed. The nucleon-nucleon 
interaction is well known from nuclear properties. But the extrapolation of 
such interactions to densities beyond nuclear saturation density is quite 
challenging. Most of the hyperon-nucleon interaction are known experimentally, 
which we use to set the hyperon-nucleon interaction
potential at saturation density. For the 
$\Lambda$, $\Sigma$ and $\Xi$ hyperons the respective potentials are 
$U_\Lambda = -28$ MeV, 
$U_\Sigma = 30$ MeV and $U_\Xi = -18$ MeV respectively. However, the coupling 
of the $\Delta$ isobars with the mesons are poorly constrained. The 
$\Delta$ isobars are commonly treated with the same coupling strengths as the 
nucleons. Studies \cite{samos,garp} based on the quark counting argument 
suggest universal 
couplings between nucleons, $\Delta$ isobars and mesons, giving the value 
of $x_{\omega\Delta}=g_{\omega\Delta}/g_{\omega N}=1$. Theoretical studies 
of Gamow-Teller transitions and M1 giant resonance in nuclei by 
Bohr and Mottelson \cite{bohr} observed a $25-40$\% reduction in transition 
strength due to the couplings to $\Delta$ isobars, indicating weaker coupling 
of the isoscalar mesons to the $\Delta$ isobars. Further, the difference 
between $x_\sigma$ and $x_\omega$ was found to be $x_\sigma-x_\omega=0.2$ in 
Hartree approximation \cite{wehr}. In view of this we scale the 
$\Delta$-coupling with the value of $x_{\omega\Delta}=0.8$. We also study 
the effect of moderate variations in the value of $x_{\omega\Delta}$ on the 
radius of neutron stars.

The paper is organized as follows: In Sec. 2, a brief outline
of the model describing the baryon structure in vacuum is discussed. 
The baryon mass is then realized by appropriately taking into account the 
center-of-mass correction, pionic correction, and gluonic correction. 
The EOS with the inclusion of the $\Delta$ isobars and the hyperons is then 
developed in Sec. 3. The results and discussions are made in 
Sec. 4. We summarize our findings in Sec. 5.

\section{Modified quark meson coupling model}

The modified quark-meson coupling model has been successful in obtaining 
various bulk properties of symmetric and asymmetric nuclear matter as well as 
hyperonic matter within the accepted constraints \cite{rnm,hss,hss2}. We now 
extend this model to include the $\Delta$ isobars 
($\Delta^-$, $\Delta^0$, $\Delta^+$, $\Delta^{++}$) along with nucleons and 
hyperons in neutron star matter under conditions of beta equilibrium and 
charge neutrality. We begin by considering baryons as composed of 
three constituent quarks  
confined inside the hadron core by a phenomenological flavor-independent 
potential, $U(r)$. Such a potential may be expressed as an admixture of equal 
scalar and vector parts in harmonic form \cite{rnm}, 
\[
U(r)=\frac{1}{2}(1+\gamma^0)V(r),
\]
with 
\begin{equation}
V(r)=(ar^2+V_0),~~~~~ ~~~ a>0. 
\label{eq:1}
\end{equation}
Here $(a,~ V_0)$ are the potential parameters. The confining interaction  
provides the zeroth-order quark dynamics of the hadron.  In the medium, the 
quark field $\psi_q({\mathbf r})$ satisfies 
the Dirac equation
\begin{equation}
[\gamma^0~(\epsilon_q-V_\omega-
\frac{1}{2} \tau_{3q}V_\rho)-{\vec \gamma}.{\vec p}
-(m_q-V_\sigma)-U(r)]\psi_q(\vec r)=0
\end{equation}
where $V_\sigma=g_\sigma^q\sigma_0$, $V_\omega=g_\omega^q\omega_0$ and
$V_\rho=g_\rho^q b_{03}$. Here $\sigma_0$, $\omega_0$, and $b_{03}$ are the
classical meson fields, and 
$g_\sigma^q$, $g_\omega^q$, and $g_\rho^q$ are the quark couplings to  
the $\sigma$, $\omega$, and $\rho$ mesons, respectively. $m_q$ is the quark
mass and $\tau_{3q}$ is the third component of the Pauli matrices. 
We can now define
\begin{equation}
\epsilon^{\prime}_q= (\epsilon_q^*-V_0/2)~~~ 
\mbox{and}~~~ m^{\prime}_q=(m_q^*+V_0/2),
\label{eprim}
\end{equation}
where the effective quark energy, 
$\epsilon_q^*=\epsilon_q-V_\omega-\frac{1}{2}\tau_{3q} V_\rho$ and 
effective quark mass, $m_q^*=m_q-V_\sigma$. We now introduce $\lambda_q$ 
and $r_{0q}$ as
\begin{equation}
(\epsilon^{\prime}_q+m^{\prime}_q)=\lambda_q~~
~~\mbox{and}~~~~r_{0q}=(a\lambda_q)^{-\frac{1}{4}}.
\label{eq:8}
\end{equation}

The ground-state quark energy can be obtained from the eigenvalue condition
\begin{equation}
(\epsilon^{\prime}_q-m^{\prime}_q)\sqrt \frac{\lambda_q}{a}=3.
\label{eq:11}
\end{equation}
The solution of equation \eqref{eq:11} for the quark  energy
$\epsilon^*_q$ immediately leads to
the mass of baryon in the medium in zeroth order as
\begin{equation}
E_B^{*0}=\sum_q~\epsilon^*_q
\label{eq:12}
\end{equation}

We next consider the spurious center-of-mass 
correction $\epsilon_{c.m.}$, the pionic correction $\delta M_{B}^\pi$ 
for restoration of chiral symmetry, and the short-distance one-gluon 
exchange contribution $(\Delta E_B)_g$ to the zeroth-order baryon 
mass in the medium. 

We have used a fixed center potential to calculate the wavefunctions of a 
quark in a baryon. To study the properties of the baryon constructed 
from these quarks, we must extract the contribution of the center-of-mass 
motion in order to obtain physically relevant results. Here, we extract 
the center of mass energy to first order in the 
difference between the fixed center and relative quark co-ordinate, 
using the method described by Guichon {\it et al.} \cite{guichon,guichon11}. 
The centre of mass correction is given by:
\begin{equation}
e_{c.m.}=e_{c.m.}^{(1)}+e_{c.m.}^{(2)},
\end{equation}
where,

\begin{equation}
e_{c.m.}^{(1)}=\sum_{i=1}^3{\left[\frac{m_{q_i}}{\sum_{k=1}^3 m_{q_k}}\frac{6}
{r_{0q_i}^2(3\epsilon'_{q_i}+m'_{q_i})}\right]}
\end{equation}

\begin{eqnarray}
e_{c.m.}^{(2)}&=&\frac{a}{2}\bigg[\frac{2}{\sum_k m_{q_k}}
\sum_im_i\langle r_i^2\rangle
+\frac{2}{\sum_k m_{q_k}}\sum_im_i\langle \gamma^0(i)r_i^2\rangle
-\frac{3}{(\sum_k m_{q_k})^2}\sum_im_i^2\langle r_i^2\rangle\nonumber\\
&-&\frac{1}{(\sum_k m_{q_k})^2}\sum_i\langle \gamma^0(1)m_i^2r_i^2\rangle-
\frac{1}{(\sum_k m_{q_k})^2}\sum_i\langle \gamma^0(2)m_i^2r_i^2\rangle\nonumber\\
&-&\frac{1}{(\sum_k m_{q_k})^2}\sum_i\langle \gamma^0(3)m_i^2r_i^2\rangle\bigg]
\end{eqnarray}
In the above, we have used for $i=(u,d,s)~$ and $k=(u,d,s)$ and the 
various quantities are defined as

\begin{equation}
\langle r_i^2\rangle = \frac{(11\epsilon_{qi}'+ m_{qi}')r^2_{0qi}}
{2(3\epsilon_{qi}'+ m_{qi}')}
\end{equation}
\begin{equation}
\langle\gamma^0(i)r_i^2\rangle=\frac{(\epsilon_{qi}'+ 11 m_{qi}')r^2_{0qi}}
{2(3\epsilon_{qi}'+ m_{qi}')}
\end{equation}
\begin{equation}
\langle\gamma^0(i)r_j^2\rangle_{i\neq j}=\frac{(\epsilon_{qi}'+ 3 m_{qi}')
\langle r^2_j\rangle}{3\epsilon_{qi}'+ m_{qi}'}
\end{equation}
\begin{table}[pt]
\renewcommand{\arraystretch}{1.4}
\setlength\tabcolsep{3pt}
\centering
\caption{\label{table0} The coefficients $a_{ij}$ and $b_{ij}$ used in the 
calculation of the color-electric and and color-magnetic energy 
contributions due to one-gluon exchange.}{
\begin{tabular}{@{}ccccccc@{}} 
\toprule
Baryon     & $a_{uu}$ & $a_{us}$ & $a_{ss}$ & $b_{uu}$ & $b_{us}$ & $b_{ss}$\\
\colrule
$N$        & -3  &  0  & 0 & 0 &  0 & 0\\
$\Delta$   &  3  &  0  & 0 & 0 &  0 & 0\\
$\Lambda$  & -3  &  0  & 0 & 1 & -2 & 1\\
$\Sigma$   &  1  & -4  & 0 & 1 & -2 & 1\\
$\Xi$      &  0  & -4  & 1 & 1 & -2 & 1\\
\botrule
\end{tabular}}
\end{table}
The pionic corrections in the model for the nucleons become
\begin{equation}
\delta M_{N}^\pi=- \frac{171}{25}I_{\pi}f_{NN\pi}^2,
\end{equation}
where, $f_{NN\pi}$ is the pseudo-vector nucleon-pion coupling constant.
Taking $w_k=(k^2+m_\pi^2)^{1/2}$,~~
$I_{\pi}$ becomes
\begin{equation}
I_{\pi}=\frac{1}{\pi{m_{\pi}}^2}\int_{0}^{\infty}dk. 
\frac{k^4u^2(k)}{w_k^2},
\end{equation}
with the axial vector nucleon form factor given as
\begin{equation}
u(k)=\Big[1-\frac{3}{2} \frac{k^2}{{\lambda}_q(5\epsilon_q^{\prime}+
7m_q^{\prime})}\Big]e^{-k^2r_0^2/4} \ .
\end{equation}
The pionic correction for $\Sigma^{0}$ and $\Lambda^{0}$ become
\begin{equation}
\delta M_{\Sigma^{0}}^{\pi}=-{\frac{12}{5}}f_{NN\pi}^2I_{\pi},
\end{equation}
\begin{equation}
\delta M_{\Lambda^{0}}^{\pi}=-{\frac{108}{25}}f_{NN\pi}^2I_{\pi}.
\end{equation}
Similarly the pionic correction for $\Sigma^{-}$ and $\Sigma^{+}$ is
\begin{equation}
\delta M_{\Sigma^{+},\Sigma^{-}}^{\pi}=-{\frac{12}{5}}f_{NN\pi}^2I_{\pi}.
\end{equation}
The pionic correction for $\Xi^{0}$ and $\Xi^{-}$ is
\begin{equation}
\delta M_{\Xi^{-},\Xi^{0}}^{\pi}=-{\frac{27}{25}}f_{NN\pi}^2I_{\pi}.
\end{equation}
For $\Delta$ baryon, the pionic correction is given by
\begin{equation}
\delta M_{\Delta}^{\pi}=-{\frac{99}{25}}f_{NN\pi}^2I_{\pi}.
\end{equation}

The one-gluon exchange interaction is provided by the interaction Lagrangian
density
\begin{equation}
{\cal L}_I^g=\sum J^{\mu a}_i(x)A_\mu^a(x) \ ,
\end{equation}
where $A_\mu^a(x)$ are the octet gluon vector-fields and $J^{\mu a}_i(x)$ is 
the $i$-th quark color current. The gluonic correction can be separated in two
pieces, namely, one from the color electric field ($E^a_i$) and another 
from the magnetic field ($B^a_i$)
generated by the $i$-th quark color current density
\begin{equation}
J^{\mu a}_i(x)=g_c\bar\psi_q(x)\gamma^\mu\lambda_i^a\psi_q(x) \ ,
\end{equation}
with $\lambda_i^a$ being the usual Gell-Mann $SU(3)$ matrices and
$\alpha_c=g_c^2/4\pi$. The contribution to the mass 
can be written as a sum of color electric and color magnetic part as
\begin{equation}
(\Delta E_B)_g=(\Delta E_B)_g^E+(\Delta E_B)_g^M \ .
\end{equation}

Finally, taking into account the specific quark flavor and spin configurations
in the ground state baryons and using the relations
$\langle\sum_a(\lambda_i^a)^2\rangle =16/3$ and
$\langle\sum_a(\lambda_i^a\lambda_j^a)\rangle_{i\ne j}=-8/3$ for
baryons, one can write the energy correction due to
color electric contribution as given in \cite{hss2}
\begin{equation}
(\Delta E_B)_g^E={\alpha_c}(b_{uu}I_{uu}^E+b_{us}I_{us}^E+b_{ss}I_{ss}^E) \ ,   
\label{enge}
\end{equation}
and due to color magnetic contributions, as
\begin{equation}
(\Delta E_B)_g^M={\alpha_c}(a_{uu}I_{uu}^M+a_{us}I_{us}^M+a_{ss}I_{ss}^M) \ ,  
\label{engm}
\end{equation}
where $a_{ij}$ and $b_{ij}$ are the numerical coefficients depending on each
baryon and are given in Table \ref{table0}. In the above, we have

\begin{eqnarray}
I_{ij}^{E}=\frac{16}{3{\sqrt \pi}}\frac{1}{R_{ij}}\Bigl[1-
\frac{\alpha_i+\alpha_j}{R_{ij}^2}+\frac{3\alpha_i\alpha_j}{R_{ij}^4}
\Bigl]
\nonumber\\
I_{ij}^{M}=\frac{256}{9{\sqrt \pi}}\frac{1}{R_{ij}^3}\frac{1}{(3\epsilon_i^{'}
+m_{i}^{'})}\frac{1}{(3\epsilon_j^{'}+m_{j}^{'})} \ ,
\end{eqnarray}
where
\begin{eqnarray}
R_{ij}^{2}&=&3\Bigl[\frac{1}{({\epsilon_i^{'}}^2-{m_i^{'}}^2)}+
\frac{1}{({\epsilon_j^{'}}^2-{m_j^{'}}^2)}\Bigl]
\nonumber\\
\alpha_i&=&\frac{1}{ (\epsilon_i^{'}+m_i^{'})(3\epsilon_i^{'}+m_{i}^{'})} \ .
\end{eqnarray}
The color electric contributions to the bare mass for nucleon and the $\Delta$ 
baryon are $(\Delta E_N)_g^{E} = 0$ and $(\Delta E_\Delta)_g^{E} = 0$. 
Therefore the one-gluon contribution for $\Delta$ becomes
\begin{equation}
(\Delta E_\Delta)_g^{M}={\frac{256\alpha_c}{3\sqrt{\pi}}}
\Big [{\frac{1}{(3{\epsilon}_u^{\prime}+m_u^{\prime})^2R_{uu}^3}}\Big ]\\
\end{equation}
The details of the gluonic correction for the nucleons and hyperons is 
given in \cite{hss2}.

Treating all energy corrections independently, the 
mass of the baryon in the medium becomes 
\begin{equation}
M_B^*=E_B^{*0}-\epsilon_{c.m.}+\delta M_B^\pi+(\Delta E_B)^E_g+
(\Delta E_B)^M_g.
\label{mass}
\end{equation}

\section{The Equation of state}
The total energy density and pressure at a 
particular baryon density, encompassing all the members of the baryon 
octet, 
for the nuclear matter in $\beta$-equilibrium can be found as
\begin{subequations}
\begin{eqnarray}
\label{engd}
{\cal E} &=&\frac{1}{2}m_\sigma^2 \sigma_0^2+\frac{1}{2}m_\omega^2 \omega^2_0
+\frac{1}{2}m_\rho^2 b^2_{03} + 
+\frac{\gamma}{2\pi^2}\sum_{B}\int ^{k_{f,B}} [k^2+{M_B^*}^2]^{1/2}k^2~dk
\nonumber\\
&+&\sum_{l}\frac{1}{\pi^2}\int_0^{k_l}[k^2+m_l^2]^{1/2}k^2dk,\\
P&=&-~\frac{1}{2}m_\sigma^2 \sigma_0^2+\frac{1}{2}m_\omega^2 \omega^2_0+
\frac{1}{2}m_\rho^2 b_{03}^2 + 
+\frac{\gamma}{6\pi^2}\sum_{B}\int ^{k_{f,B}} \frac{k^4~ dk}
{[k^2+{M_B^*}^2]^{1/2}} \nonumber\\
&+& \frac{1}{3}\sum_{l}\frac{1}{\pi^2}\int_0^{k_l}\frac{k^4dk}{[k^2+m_l^2]^{1/2}},
\end{eqnarray} 
\end{subequations}
where $\gamma$ is the spin degeneracy factor for nuclear matter. 
For the nucleons and hyperons $\gamma=2$ and for the $\Delta$ baryons 
$\gamma=4$. Here  
$B=N,~\Delta,~\Lambda,~\Sigma^{\pm},~\Sigma^0,~\Xi^-,~\Xi^0$ and $l=e,\mu$.

The chemical potentials, necessary to define the $\beta-$ equilibrium 
conditions, are given by
\begin{equation}
\mu_B=\sqrt{k_B^2+{M_B^*}^2}+g_\omega\omega_0+g_\rho\tau_{3B}b_{03}
\end{equation}
where $\tau_{3B}$ is the isospsin projection of the baryon B.

The lepton Fermi momenta are the positive real solutions of
$(k_e^2 + m_e^2)^{1/2} =  \mu_e$ and
$(k_\mu^2 + m_\mu^2)^{1/2} = \mu_\mu$. The equilibrium composition
of the star is obtained by solving the equations of motion of meson fields in 
conjunction with the charge neutrality condition, given in 
Eq. (\ref{neutral}), 
at a given total baryonic density $\rho = \sum_B \gamma k_B^3/(6\pi^2)$. 
The effective masses of the baryons are
obtained self-consistently in this model.

Since the neutron star time scale is quite long we need to consider the 
occurence of weak processes in its matter. Moreover, for stars in which the 
strongly interacting particles are baryons, the
composition is determined by the requirements of charge neutrality
and $\beta$-equilibrium conditions under the weak processes
$B_1 \to B_2 + l + {\overline \nu}_l$ and $B_2 + l \to B_1 + \nu_l$.
After de-leptonization, the charge neutrality condition yields
\begin{equation}
q_{\rm tot} = \sum_B q_B \frac{\gamma k_B^3}{6\pi^2}
+ \sum_{l=e,\mu} q_l \frac{k_l^3}{3\pi^2}  = 0 ~,
\label{neutral}
\end{equation}
where $q_B$ corresponds to the electric charge of baryon species $B$
and $q_l$ corresponds to the electric charge of lepton species $l$. Since
the time scale of a star is effectively infinite compared to the weak
interaction time scale, weak interaction violates strangeness conservation.
The strangeness quantum number is therefore not conserved
in a star and the net strangeness is determined by the condition of
$\beta$-equilibrium which for baryon $B$ is then given by
$\mu_B = b_B\mu_n - q_B\mu_e$, where $\mu_B$ is the chemical potential
of baryon $B$ and $b_B$ its baryon number. Thus the chemical potential of any
baryon can be obtained from the two independent chemical potentials $\mu_n$
and $\mu_e$ of neutron and electron respectively.

In the present work, the baryon couplings are given by,   
$$g_{\sigma B}=x_{\sigma B}~ g_{\sigma N},~~g_{\omega B}=x_{\omega B}~ 
g_{\omega N}, ~~g_{\rho B}=x_{\rho B}~ g_{\rho N},$$
where $x_{\sigma B}$, $x_{\omega B}$ and $x_{\rho B}$ are equal to $1$ for the
nucleons and acquire different values in different parameterisations for the
other baryons. Information about the hyperon couplings can be obtained from the 
levels in $\Lambda$ hyper-nuclei \cite{chrien}. 
We note that the $s$-quark is unaffected by the $\sigma$- and $\omega$-
mesons i.e. $g_\sigma^s=g_\omega^s=0$. The coupling of the $\Delta$ resonances 
are constrained poorly due to their unstable nature. Earlier works 
\cite{samos,garp} based on the quark counting argument considered simple 
universal choice of couplings of the $\Delta$ with the mesons. Wehrberger 
{\it et al.} \cite{wehr} carried out studies of $\Delta-$baryon excitation 
in finite nuclei in linear Walecka model and reproduced properties of some 
finite nucleus. They constrained the scaling to 
$0 \lesssim x_{\sigma\Delta}-x_{\omega\Delta}\lesssim 0.2$.

The vector mean-fields $\omega_0$ and $b_{03}$ are determined through
\begin{equation}
\omega_0=\frac{g_\omega}{{m_\omega}^2} \sum_B x_{\omega B}\rho_B~~~~~~~~~~
b_{03}=\frac{g_\rho}{{2m_\rho}^2} \sum_B x_{\rho B}\tau_{3B}\rho_B,
\label{omg}
\end{equation}
where $g_\omega=3 g_\omega^q$ and $g_\rho= g_\rho^q$.
Finally, the scalar mean-field $\sigma_0$ is fixed by
\begin{equation}
\frac{\partial {\cal E }}{\partial \sigma_0}=0.
\label{sig}
\end{equation}
The iso-scalar scalar and iso-scalar vector couplings $g_\sigma^q$ and $g_\omega$
are fitted to the saturation density and binding energy for nuclear
matter. The iso-vector vector coupling $g_\rho$ is set by fixing 
the symmetry energy at $J=32.0$ MeV. 
For a given baryon density, $\omega_0$, $b_{03}$, and $\sigma_0$ are
calculated from Eq. \eqref{omg} and \eqref{sig}, respectively. 

The relation between the mass and 
radius of a star with its central energy density can be obtained by integrating 
the Tolman-Oppenheimer-Volkoff (TOV) equations \cite{tov,tov11} given by,
\begin{equation}
\frac{dP}{dr}=-\frac{G}{r}\frac{\left[{\cal E}+P\right ]\left[M+
4\pi r^3 P\right ]}{(r-2 GM)},
\label{tov1}
\end{equation}
\begin{equation}
\frac{dM}{dr}= 4\pi r^2 {\cal E} ,
\label{tov2}
\end{equation}
with $G$ as the gravitational constant and $M(r)$ as the enclosed gravitational
mass. We have used $c=1$. Given an EOS, these equations can be integrated 
from the origin as an initial value problem for a given choice of the 
central energy density, $(\varepsilon_0)$. Of particular importance is the 
maximum mass obtained from and the 
solution of the TOV equations. 
The value of $r~(=R)$, where the pressure vanishes defines the
surface of the star. The surface gravitational redshift $Z_s$ is defined as,
\begin{equation} 
Z_s=\left(1-\frac{2GM}{R}\right)^{-1/2}-1
\end{equation}

\section{Results and Discussion}
The MQMC model has two potential parameters, $a$ and  $V_0$ which are obtained 
by fitting the nucleon mass $M_N=939$ MeV and charge radius of the 
proton $\langle r_N\rangle=0.87$ fm in free space. Keeping the value of the 
potential parameter $a$ same as that for nucleons, we obtain $V_0$ for the 
$\Lambda$, $\Delta$, $\Sigma$ and $\Xi$ baryons by fitting their 
respective masses 
to $M_{\Lambda}=1115.6$ MeV, $M_{\Delta}=1232$ MeV, $M_{\Sigma}=1193.1$ MeV and 
$M_{\Xi}=1321.3$ MeV. 
The set of potential parameters for the baryons along with their respective 
energy corrections at zero density are given in Table \ref{table1}. 
\begin{table}[t]
\centering
\renewcommand{\arraystretch}{1.4}
\setlength\tabcolsep{3pt}
\caption{\label{table1} The potential parameter $V_0$ obtained for the
quark mass $m_u=m_d=150$ MeV, $m_s=300$ MeV with $a=0.69655$~fm$^{-3}$.}
{\begin{tabular}{@{}ccc@{}}
\toprule
Baryon     & $M_{B}$(MeV) & $V_0$(MeV) \\
\colrule
$N$        & 939    &  44.05  \\
$\Delta$   & 1232   &  102.40 \\
$\Lambda$  & 1115.6 &  50.06  \\
$\Sigma$   & 1193.1 &  66.44  \\
$\Xi$      & 1321.3 &  66.82  \\
\botrule
\end{tabular}}
\end{table}
%

\begin{table*}[ht]
\centering
\renewcommand{\arraystretch}{1.4}
\setlength\tabcolsep{3pt}
\caption{\label{table2}Parameters for nuclear matter. They are determined
from the binding energy per nucleon, $E_{B.E}=B_0 \equiv{\cal E} /\rho_B - M_N 
= -15.7$~MeV and pressure, $P=0$ at saturation density
$\rho_B=\rho_0=0.15$~fm$^{-3}$. Also shown are the values of the nuclear 
matter incompressibility $K$ and the slope of the symmetry energy $L$ 
for the quark mass $m_q=150$ MeV.}
{\begin{tabular}{@{}ccccccc@{}}
\toprule
$m_q$ & $g^q_\sigma$ & $g_\omega$ & $g_\rho$ & $M_N^*/M_N$ & K 
& L\\

(MeV)& & & & & (MeV)& (MeV)\\
\colrule
150   &4.39952  &6.74299 &8.79976 &0.87 &292 &86.39\\
\botrule
\end{tabular}}

\end{table*}
\begin{figure}
\centering
\includegraphics[width=8.cm,angle=0]{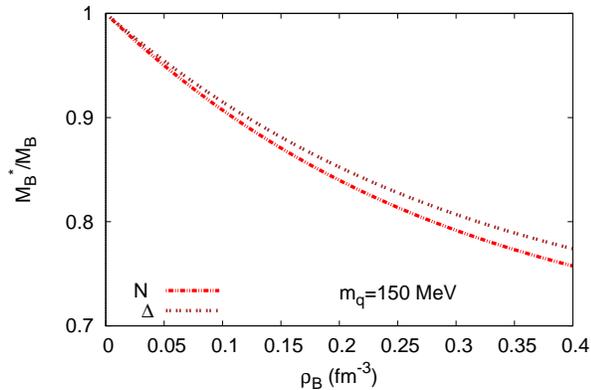}
\caption{Effective baryon mass as a function of baryon density.}
\label{fig1}
\end{figure}
\begin{figure}
\centering
\includegraphics[width=8.cm,angle=0]{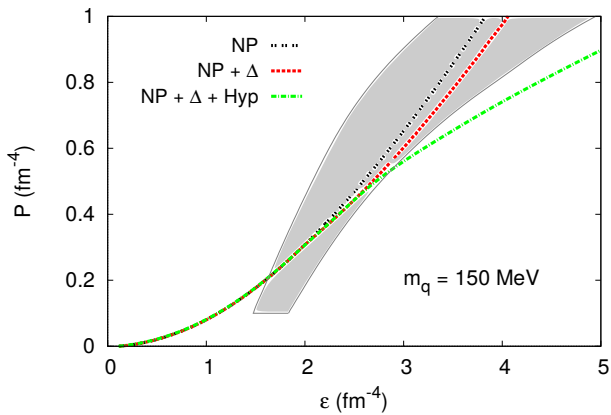}
\caption{Total pressure as a function of the energy density for various 
composition of the stellar matter at quark mass $m_q=150$ MeV. 
The shaded region shows the empirical EOS obtained by Steiner {\it et al} 
from a heterogeneous set of seven neutron stars.}
\label{fig2}
\end{figure}
%
\begin{figure}
\centering
\includegraphics[width=8.cm,angle=0]{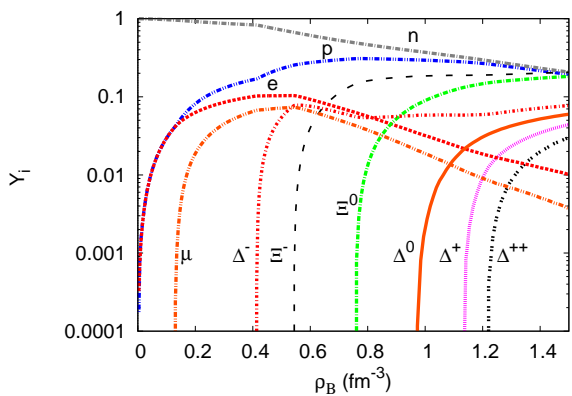}
\caption{Particle fraction as a function of the baryon density indicating the 
onset of the $\Delta$ isobars at quark mass $m_q=150$ MeV and 
$x_{\omega\Delta}=0.8$.}
\label{fig3}
\end{figure}
\begin{figure}
\centering
\includegraphics[width=8.cm,angle=0]{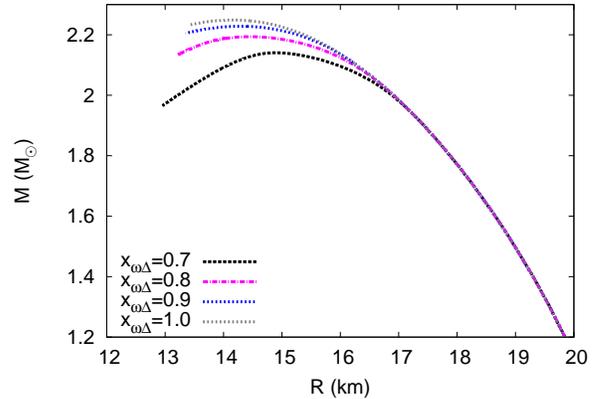}
\caption{Gravitational mass as a function of radius for 
various couplings of the $\Delta$ isobars at quark mass $m_q=150$ MeV.}
\label{fig5}
\end{figure}

The quark meson couplings $g_\sigma^q$, $g_\omega=3g_\omega^q$, 
and $g_\rho=g_\rho^q$ are fitted self-consistently for the nucleons to obtain 
the correct saturation properties of nuclear matter binding energy, 
$E_{B.E.}\equiv B_0={\cal E}/\rho_B-M_N=-15.7$ MeV, pressure, $P=0$, and 
symmetry energy $J=32.0$ MeV at 
$\rho_B=\rho_0=0.15$ fm$^{-3}$. 

We have taken the standard values for the 
meson masses; namely, $m_\sigma=550$ MeV, $m_\omega=783$ MeV and 
$m_\rho=763$ MeV. The values of the quark meson couplings, 
$g_\sigma^q$, $g_\omega$, and $g_\rho$ at quark mass $150$ MeV 
is given in Table \ref{table2}.

The couplings of the hyperons to the $\sigma$-meson need not be fixed 
since we determine the effective masses of the hyperons self-consistently. 
The hyperon couplings to the $\omega$-meson 
are fixed by determining $x_{{\omega}B}$. 
The value of $x_{{\omega}B}$ is obtained from the hyperon potentials in nuclear 
matter, $U_B=-(M_B^*-M_B)+x_{{\omega}B}g_{\omega}\omega_0$ for $B={\Lambda},{\Sigma}$ 
and $\Xi$ as $-28$ MeV, $30$ MeV and $-18$ MeV respectively. 
For the quark masses $150$ MeV the corresponding values 
for $x_{{\omega}B}$ are given in 
Table \ref{table3}. The $\Delta$-coupling to the $\omega$-meson is fixed 
at $x_{{\omega}\Delta} = 0.8$. The value of $x_{{\rho}B}=1$ is fixed for 
all baryons. 

\begin{table*}[ht]
\centering
\renewcommand{\arraystretch}{1.4}
\setlength\tabcolsep{3pt}
\caption{\label{table3}$x_{{\omega}B}$ determined by fixing the 
 potentials for the hyperons.}
{\begin{tabular}{@{}cccc@{}}
\toprule
$m_q$ (MeV) & $x_{{\omega}\Lambda}$& $x_{{\omega}\Sigma}$&
$x_{{\omega}\Xi}$\\ 
& $U_\Lambda=-28$ MeV& $U_\Sigma=30$ MeV& $U_\Xi=-18$ MeV\\
\colrule
150     &0.81309  & 1.58607 & 0.24769 \\
\botrule
\end{tabular}}

\end{table*}

The $\Lambda$ hyperon potential has been chosen from the measured single  
particle  levels of $\Lambda$ hypernuclei from mass numbers $A = 3$ to $209$ 
\cite{millener,millener1} of the binding of $\Lambda$ to symmetric 
nuclear matter. 
Studies of $\Sigma$ nuclear interaction \cite{mares,bart} from the analysis 
of $\Sigma^-$ atomic data indicate a repulsive isoscalar potential in the 
interior of nuclei. 
Measurements of the final state interaction of $\Xi$ hyperons produced in 
($K^-,K^+$) reaction on $^{12}C$ in E224 experiment at KEK \cite{fukuda} and 
E885 experiment at AGS \cite{khaustov} indicate a shallow attractive 
potential $U_\Xi \sim -16$ MeV and $U_\Xi \sim-14$ or less respectively. 
In view of this we consider the $\Xi$ hyperon potential at $U_\Xi = -18$ MeV. 
Fig. \ref{fig1} shows the effective mass of the nucleons and $\Delta$. 
With increasing density the effective mass decreases due to the attractive 
$\sigma$ field for the baryons. 

The EOS for different compositions of neutron star 
matter is shown in Fig. \ref{fig2}. It is observed that with the inclusion of 
$\Delta$, the EOS becomes softer than for matter containing only the nucleons. 
For matter containing the nucleons, delta and the hyperons, we observe 
significant decrease of stiffness. Infact, when both the hyperons and the 
$\Delta$ baryons are present, the softness appears at a density of 
$\rho_B=0.41$~fm$^{-3}$, which is lower than the density of 
$\rho_B=0.54$~fm$^{-3}$, when the softness increases for matter containing 
nucleons and $\Delta$'s. The shaded region shows the empirical EOS obtained by
Steiner {\it et al.} from a heterogeneous set of seven neutron stars with well
determined distances \cite{awsteiner}.

The composition of the matter is shown in Fig. \ref{fig3}.  
which shows the particle fractions for $\beta$-equilibriated 
matter. At densities below the saturation value the $\beta$-decay of 
neutrons to muons are allowed and thus muons start to populate. 
At higher densities the lepton fraction begins to fall since 
charge neutrality can now be maintained more economically with the 
appearance of negative baryon species. Since the $\Delta^-$ can replace the 
neutron and electron at the top of the Fermi sea, it appears first at a 
density of $\rho_B=0.41$~fm$^{-3}$. This is 
followed by the appearance of ${\Xi}^-$. The sequence of appearance of the 
$\Delta$ resonances is consistent with the notion of charge-favored or 
unfavored species \cite{glend}. As such, the first $\Delta$ species to appear 
is $\Delta^-$, followed by the $\Delta^0$, $\Delta^+$ and $\Delta^{++}$.
The slope of the symmetry energy $L$ also plays a key role in the 
appearance of $\Delta$ resonances. Drago {\it et al.} 
\cite{pigato} constraining $L$ in the 
range $40<L<62$ MeV have observed the appearance of $\Delta$ close to twice 
the saturation density.
At high densities all baryons tend to saturate. 
Moreover, the $\Sigma$ hyperon is not present in the matter distribution 
for the given set of potentials since we have chosen a repulsive potential 
for it.  

\begin{table}[t]
\centering
\caption{\label{table8}Mass-radius relationaship of neutron stars for different
vector coupling strength of the $\Delta$ isobars at $m_q=150$ MeV.}
{\begin{tabular}{@{}ccc@{}}
\toprule
$x_{\omega\Delta}$&  M$_{max}$ &  R \\
  & ($M_{\odot}$) &  (km) \\
\colrule
0.70       & 2.14   & 14.88 \\
0.80       & 2.19   & 14.40 \\
0.90       & 2.22   & 14.28 \\
1.00       & 2.24   & 14.15 \\
\botrule
\end{tabular}}

\end{table}
\begin{table}[t]
\centering
\caption{\label{table9}Stellar properties obtained at different compositions of the 
star matter for quark mass $m_q=150$ MeV.}
{\begin{tabular}{@{}cccc@{}}
\toprule
$m_q$  &  Composition      &  M$_{max}$     &  R \\
(MeV)  &                   &  ($M_{\odot}$) & (km) \\
\colrule
150    & NP                & 2.25   & 14.0 \\
       & NP+$\Delta$       & 2.19   & 14.4 \\
       & NP+$\Delta$+HYP   & 2.15   & 15.4 \\
\botrule
\end{tabular}}

\end{table}   
Since the vector coupling of the $\Delta$ are not constrained 
by the properties of saturated nuclear matter, we study the effect of moderate 
variations in the strength of the vector coupling of the $\Delta$ on the 
mass-radius of the neutron star. Considering only the nucleon and $\Delta$ 
composition of the matter, we plot in Fig. \ref{fig5} the gravitational mass 
as a function of radius by changing the coupling strength $x_{\omega\Delta}$ 
of the $\Delta$ isobars. By decreasing the coupling strength from 
$x_{\omega\Delta}=1.0$ to $x_{\omega\Delta}=0.7$, we observe a gradual 
decrease in the maximum mass of the star, see Table \ref{table8}. 
This follows from the fact that 
by decreasing the interaction strength of the $\Delta$  with respect to 
the nucleons, the EOS becomes softer with a consequent decrease in the 
maximum mass of the star. We also observe a decrease in the radius with 
increasing coupling strength.

\begin{figure}
\centering
\includegraphics[width=8.cm,angle=0]{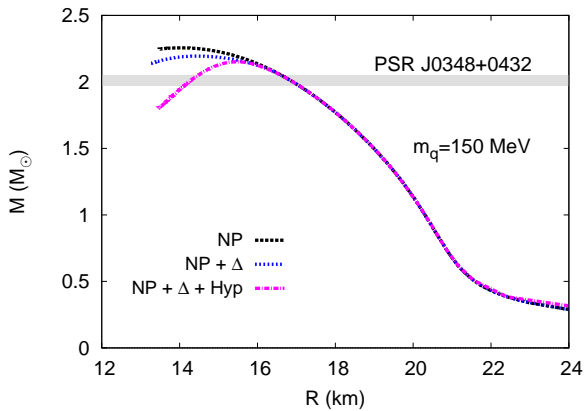}
\caption{Gravitational mass as a function of radius for varying composition 
of star matter at quark mass $m_q=150$ MeV and $x_{\omega\Delta}=0.8$.}
\label{fig4}
\end{figure}

In Fig. \ref{fig4} we plot the mass-radius relations for the three possible 
compositions of neutron star matter at $m_q=150$ MeV. A stiffer EOS 
corresponding to matter with nucleons only gives the maximum star mass 
of $M_{star}=2.25 M_{\odot}$. With the appearance of the $\Delta$ isobars, 
mass decreases by $0.06 M_{\odot}$ to $M_{star}=2.19 M_{\odot}$. The inclusion 
of the hyperons further softens the EOS resulting in a corresponding decrease 
in the maximum mass to $M_{star}=2.15 M_{\odot}$. 
The results are shown 
in Table \ref{table9}. The recently observed pulsar PSR J0348+0432 provide 
a mass constraint of $2.01\pm0.04 M_{\odot}$ \cite{antoniadis} while an 
earlier accurately measured pulsar PSR J1614-2230 gives a mass 
of $1.97\pm0.04 M_{\odot}$ \cite{demorest}. From our calculations we obtain a 
range of masses varying from $2.15 M_{\odot}$ to $2.25 M_{\odot}$ depending 
on the composition of the matter. 
Though in the present model we are able to meet the mass constraint,  
we do not get a lower radius. Such problems are also found in many other 
models. One of the way out in such model based calculations is to consider a 
compact star with mixed phases of hadrons and quarks. At present we consider 
only a hadronic phase. The work in this regard to meet the compactness 
in the present model is in progress. 

\begin{figure}
\centering
\includegraphics[width=8.cm,angle=0]{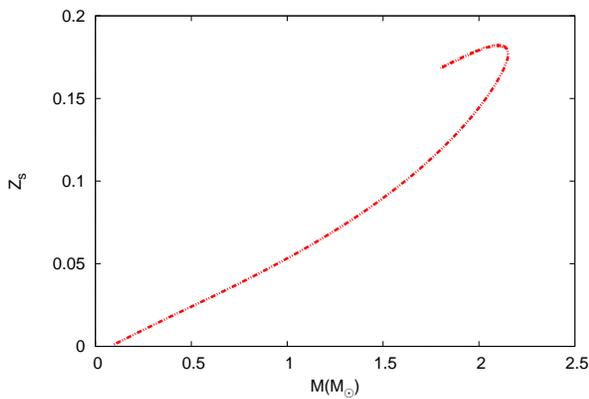}
\caption{Surface gravitational redshift as a function of star mass at quark 
mass $m_q=150$ MeV and $x_{\omega\Delta}=0.8$.}
\label{fig6}
\end{figure}

Fig. \ref{fig6} shows the gravitational redshift versus the gravitational mass 
of the neutron star at quark mass $m_q=150$ MeV and $x_{\omega\Delta}=0.8$. It 
also shows the maximum redshift (redshift corresponding to the maximum mass) 
which, for the present work comes out to be $Z_s^{max}=0.17$. This is well 
below the upper bound on the surface redshift for subluminal equation 
of states, i.e. $z_s^{CL}=0.8509$ \cite{haensel}.    

\section{Conclusion}

In the present work we have studied the possibility of the appearance the 
$\Delta$ isobars in dense matter relevant to neutron stars. We have developed 
the EOS using a modified quark-meson coupling model which considers the baryons to be composed of three independent relativistic quarks confined by an equal 
admixture of a scalar-vector harmonic potential in a background of scalar and 
vector mean fields. Corrections to the centre of mass motion, pionic and 
gluonic exchanges within the nucleon are calculated to obtain the 
effective mass of the baryon. 
The baryon-baryon interactions are realised by the quark coupling to the 
$\sigma$, $\omega$ and $\rho$ mesons through a mean field approximation. The 
nuclear matter incompressibility $K$ is determined to agree with experimental 
studies. Further, the slope of the nuclear symmetry is calculated which also 
agrees well with experimental observations. 

By varying the composition of the matter we observe the variation in the 
degree of stiffness of the EOS and the corresponding effect on the maximum 
mass of the star. As predicted theoretically, we observe that the inclusion 
of the $\Delta$ and hyperon degrees of freedom softens the EOS and hence 
lowers the maximum mass of the neutron star. The so called $\Delta$ and 
hyperon puzzles state that the presence of the $\Delta$ isobars and hyperons 
would decrease the maximum star mass below the recently observed masses 
of the pulsars PSR J0348+0432 and PSR J1614-2230. In the present work, we are 
able to achieve the observed mass constraint and at the same time 
satisfy the theoretical predictions of the possibility of existence of 
higher mass baryons in highly dense matter. Moreover, we study the effect of 
moderate variations in the strength of the vector coupling of the $\Delta$ 
resonances and observe a decrease in the radius of neutron stars with an 
increase in the coupling strength.
\section*{ACKNOWLEDGMENTS}
The authors would like to acknowledge the financial assistance from 
BRNS, India for the Project No. 2013/37P/66/BRNS.

\end{document}